\documentstyle[12pt]{article}

\def\tou#1{{\lower1.2ex\hbox{$\longrightarrow$}\atop
        {\lower-.7ex\hbox{$\scriptscriptstyle #1 $}}}}
\def\lsim{{\lower1.2ex\hbox{$<$}\atop
        {\lower-.7ex\hbox{$\sim$}}}}
\def\gsim{{\lower1.2ex\hbox{$>$}\atop
        {\lower-.7ex\hbox{$\sim$}}}}

\begin{document}

\begin{titlepage}
\rightline {TKYM-93-2\  \  \  \   }
\rightline {Si-93-8 \  \  \  \   }

\vskip 3truecm
\centerline {\Large\bf  On the Segregation Phenomenon in}
\vskip 0.3truecm
\centerline {\Large\bf Complex Langevin Simulation\footnote{
  Work supported in part by the Deutsche Forschungsgemeinschaft,
  Az.~Schu95/7-2}}

\vskip 2truecm
\centerline{\bf K. Fujimura$^*$, K. Okano$^*$, L. Sch\"ulke, K. Yamagishi$^*$
and B. Zheng
}

\medskip

\centerline{FB Physik, Universit\"at-Gesamthochschule Siegen, D-57068
Siegen, Germany}

\centerline{$^*$ Tokuyama University, Tokuyama-shi,
Yamaguchi, Japan 745}

\vskip 3cm

\abstract

In the numerical simulation of certain field theoretical models,
the complex Langevin simulation has been believed to fail due to
the violation of ergodicity.
We give a detailed analysis of this problem
based on a toy model with one degree of freedom ($S=-\beta\cos\theta$).
We find that the failure is not due to the
defect of complex Langevin simulation itself, but rather
to the way how one treats the singularity appearing in the drift force.
An effective algorithm is proposed by which one can simulate
the ${1/\beta}$ behaviour of the expectation value
$<\cos\theta>$ in the small $\beta$ limit.

\vskip .5truecm
\end{titlepage}

\section{Introduction}

\setcounter{equation}{0}

The complex Langevin simulation is a challenging idea for
quantum systems with complex actions
\cite{PAR83,KLA83}, where the conventional
Monte Carlo methods are not efficient.
It has been applied to a variety of physical systems such as
QCD at finite temperature and chemical potential
\cite{KAR85,ILG86,BIL87,BIL88}, gauge theories
with  external charges or fermions
\cite{AMB85,AMB86,GAU86,KLA92}. Big successes could be
obtained in case the complex Langevin simulation works.
In some cases, however, it fails to give correct
results.

One typical reason for the failure of the complex Langevin simulation
are instabilities, where blow-up solutions are observed due to
the non-positive definite Fokker-Planck Hamiltonian
\cite{KLA85,AMB85}.
To tackle this problem, the role of a kernel is investigated and some
progress has been made \cite{OKAM89,OKA91,SOD88}.
In principle, if a suitable
kernel is introduced, blow-up solutions could be avoided and
correct results may be obtained. Recently some more discussions
about the convergence of the complex Langevin equation
\cite{GAU92,LEE93}
and its spurious solutions \cite{SAL93} have been presented.

In this paper another type of failure is concerned.
In the pratical simulation of certain models,
no blow-up solutions are observed but the results are not correct.
One important example is the non-Abelian gauge theory
coupled to external charges \cite{AMB86,FLO86}. The violation of
ergodicity due to the so-called segregation theorem
has been blamed for this failure \cite{FLO86}.
This defect may not be remedied by the introduction of
a kernel (at least up to now no suitable one has been found)
\cite{SAL93}.
It has been pointed out that the singular
points corresponding to the zeroes in the probability
distribution may play an important role in this problem \cite{SCH91}.

The purpose of this paper is to present a detailed analysis of
the latter problem above. In the next section a brief
review in the context of a simplified model with one-degree
of freedom is given. Our new observation to the failure
and some theoretical supports are discussed in section $3$ and
$4$. In sections $5$ and $6$ the possible algorithms
for simulating the $\delta$-type singular drift forces
are presented. Finally the conclusions follow.

\section{Brief review of the problem}

\setcounter{equation}{0}

\subsection{The idea and its failure}

Let us consider the following simple integral,
        \begin{equation}
<\cos\theta>=
    \frac{\int_0^{2\pi} d\theta\cos^2\theta\exp\{\beta\cos\theta\}}
         {\int_0^{2\pi} d\theta\cos  \theta\exp\{\beta\cos\theta\}}
         =\frac{I_0(\beta)}{I_1(\beta)}-\frac{1}{\beta}\;.
        \label{10}
        \end{equation}
The analytical result shows that the quantity is singular
$(\sim 1/\beta)$ in the limit $\beta \to 0$.
In this limit the denominator becomes very small, therefore, to simulate
the quantity to high precision
by calculating numerator and denominator separately
becomes very difficult.
In order to overcome this difficulty, an idea is to apply the Langevin method.
For this purpose one  first rewrites the integral as
        \begin{equation}
<\cos\theta>=
        \int_0^{2\pi} d\theta \cos\theta P_{eff}(\theta)=
        \frac{\int_0^{2\pi} d\theta\cos\theta\exp\{-S_{eff}\}}
         {\int_0^{2\pi} d\theta \exp\{-S_{eff}\}}\;,
        \label{20}
        \end{equation}
where
        \begin{equation}
P_{eff} \propto e^{-S_{eff}}\;,\qquad
S_{eff}=-\beta\cos\theta - \ln(\cos\theta).
        \label{30}
        \end{equation}
This action is complex for $\cos\theta<0$. As the
distribution $P_{eff}\propto e^{-S_{eff}}$ is not positive definite, usual
Monte Carlo method can not be applied. On the other hand the Langevin
equation

        \begin{equation}
\dot \theta=-\beta\sin\theta -\tan\theta +\eta
        \label{40}
        \end{equation}
can in principle be solved and may be used for the simulation of (\ref{20}).
Unfortunately, this Langevin equation turns out to fail
completely to generate the desired configurations for $\beta\lsim2.5$, and
therefore cannot give the correct $1/\beta$ behaviour. This can be seen
in Fig.~\ref{F1}.

        \begin{figure}[p]
        \vspace*{5cm}
        \caption{}
         \label{F1}
         \begin{quotation}
$<\cos\theta>$ simulated by the complex Langevin equation compared with
the theoretical curve (solid line) given by (\ref{10}). The dashed line and
the dot-dashed line corresponds, respectively, to the theoretical prediction
$<\cos\theta>_1$ and $<\cos\theta>_2$ explained in subsection 2.2.
The dotted line corresponds to  $<\cos\theta>_{abs}$ explained in section 3.
         \end{quotation}
         \end{figure}

\subsection{Conventional interpretation of the failure \newline
                 and the segregation phenomenon}

In order to give the $1/\beta$ singularity,
the solution of the Langevin equation
$\theta=\theta_r+i\theta_i$ should get a big imaginary part. This is
clear because $\Re\cos\theta=\cos\theta_r\cosh\theta_i$ and
$\cos\theta_r\leq 1$. Contrary to this expectation, an important
observation given in \cite{AMB86a,FLO86} is that the configurations
obtained through the updations by use of the Langevin
equation (\ref{40}) are almost real. This phenomenon is always observed
independently of which initial configuration one takes, and was called
``the collapse of the complex distribution to the real distribution''.

This is surely the direct reason why the Langevin simulation fails
to give the ${1/\beta}$-singularity at $\beta\to 0$ as is shown in
Fig.~\ref{F1}.

In addition, in \cite{FLO86} the segregation theorem has been applied to
explain the wrong result obtained by the Langevin simulation.
Due to the collapse of the complex
distribution to a real one we can restrict the discussion to the
stochastic process on the real axis. Let us assume that there exists a
domain $D$ on the real axis with a boundary on which the probability
distribution vanishes. In this case, the segregation theorem states that
the probability that a real diffusion process starting inside $D$ exits the
domain or stays permanently near the boundary of the domain, is
zero. In case of our model (\ref{30}), the zeros of $P_{eff}$
 divide the total configuration
space ${\cal D}\equiv [0,2\pi]$ (note that we will adopt periodic
boundary conditions) into two subdomains: $D_{1}\equiv [0, {\pi \over 2})
\oplus ({3\pi \over 2}, 2\pi]$ and $D_{2}\equiv ({\pi \over 2}, {3\pi \over
2})$. If the segregation theorem is applied, the Langevin simulation should
give one of the following results,

        \begin{equation}
<\cos\theta>_i \equiv \frac{\int_{D_i} \cos\theta e^{-S_{eff}}}
                    {\int_{D_i} e^{-S_{eff}}},\qquad i=1 {\  or\ } 2.
        \label{50}
        \end{equation}
depending on the starting point of the simulation.

Note that the the above integrals have the same form as (\ref{20}), while the
integration regions are restricted to the subdomain
$D_{1}$ or $D_{2}$. These results have also been plotted in
Fig.~\ref{F1} by dotted and dot-dashed lines. In this figure one may see
the clear discrepancy between the result of the Langevin simulation and the
above prediction from the segregation phenomenon. The situation was quite
different in \cite{FLO86}. In that paper the simulated result and the
prediction given by the segregation assumption (especially
$<\cos\theta>_1$) agree quite well, by which the authors concluded
that the failure of their simulation is connected to the segregation
phenomenon or, in other words, the violation of the ergodicity.

\section{A new observation}

\setcounter{equation}{0}

\subsection{It is not the segregation but!}

Before discussing the difference between the result of the Langevin
simulation obtained here and that of \cite{FLO86}, let us first describe
the algorithm used in more detail.

In numerically solving the Langevin
equation, it sometimes happens that the configuration comes near to the
singular points where the drift force becomes very big. To get an accurate
numerical solution in such a case, one should adjust the fictitious time
step. Here we have followed the way used in \cite{FLO86}. Let us write the
discretized Langevin equation as $\Delta\theta=D(\theta)\Delta t
+\sqrt{2\Delta t}\xi$ $(<\xi^2>=1)$. The criterion to get an accurate
numerical solution is represented by $|D(\theta)\Delta t| \ll
|\sqrt{2\Delta t}\xi |$, which gives $\Delta t \ll \frac{2}{D^2(\theta)}$ (
we have used $\xi^2\sim 1$). In discretizing the Langevin equation, we
normally use a certain prefixed fictitious time step $\Delta t=\Delta t_p$.
In case that the configuration comes near to the singular points and
therefore the relation $\Delta t_p < \frac{2}{D^2(\theta)}$ is not satisfied we
replace $\Delta t_p$ by $\Delta t = \frac{2}{D^2(\theta)}$. Repeating
the updations by the use of this adjusted time step until $\sum
\Delta t=\Delta t_p$, we include the last configuration in
the set of ensemble elements.

Let us come back to the discussion about the difference of our Langevin
result in Fig.~\ref{F1} and that in \cite{FLO86}. As was mentioned above
the algorithm used here to update the configuration is almost the same as
that used in \cite{FLO86}. The big difference in the result, however,
arises from the number of iterations taken in solving the Langevin
equation. In order to show this we plot in Fig.~\ref{F2} the result of
$<\cos\theta>$ versus the number of iterations.

        \begin{figure}[p]
        \vspace*{5cm}
        \caption{}
         \label{F2}
         \begin{quotation}
$<\cos\theta>$ simulated by the complex Langevin equation versus the
number of iterations. Dotted line, dashed line and dot-dashed line
corresponds respectively to $<\cos\theta>_{abs}$,
$<\cos\theta>_1$ and $<\cos\theta>_2$.
         \end{quotation}
         \end{figure}

When one calculates the average from less than $500$ iterations, the result
coincides with $<\cos\theta>_i$, ($i=1$ or $i=2$) in (\ref{50}), given by
the segregation assumption. Which value ($<\cos\theta>_1$ or
$<\cos\theta>_2$) will result from the simulation depends in
which region, $D_1$ or $D_2$, the
starting point of the iteration was chosen.
calculates the same quantity from a much bigger ensemble,
the result converges to a different value that does not depend on
the choice of the initial configuration. See Fig.~\ref{F1}, which is
clearly showing that the segregation is observed only within those
configurations in which the number of updations is too small.

This can much more directly be seen in Fig.~\ref{F3}, where the probability
distribution within an ensemble of, respectively, $300$, $1000$
and $30000$ configurations are shown. In Fig.~\ref{F3}a, the configurations
are distributed in only one of the regions $D_1$ and $D_2$. On the other
hand, in Fig.~\ref{F3}b and in Fig.~\ref{F3}c the distribution of the
configuration covers the whole region of ${\cal D}$.

        \begin{figure}[p]
        \vspace*{5cm}
        \caption{}
         \label{F3}
         \begin{quotation}
Probability distribution within an ensemble of configurations.
The figure (a),(b) and (c), respectively, is that of ensembles
obtained by $300$, $1000$ and $30000$
iterations using the Langevin equation (\ref{40}).
A line plotted by circles represents the distribution $P_{abs}\propto
e^{-S_{abs}}$,
while the solid line shows the theoretical prediction from the segregation
assumption
in $D_1$. The double-circle is the starting point of the iterative solution.
         \end{quotation}
         \end{figure}

Within the above two results, another important point needs to be stressed.
It is the fact that the simulated results in Fig.~\ref{F2} coincide
exactly with\cite{SCH91}

        \begin{equation}
<\cos\theta>_{abs}=
        \int_0^{2\pi} d\theta \cos\theta P_{abs}(\theta)=
        \frac{\int_0^{2\pi} d\theta\cos\theta \exp\{-S_{abs}\}}
         {\int_0^{2\pi} d\theta \exp\{-S_{abs}\}}\;,
        \label{60}
        \end{equation}
where
        \begin{equation}
P_{abs} \propto e^{-S_{abs}}\;,\qquad
S_{abs}=-\beta\cos\theta - \ln|\cos\theta|.
        \label{70}
        \end{equation}
Also note that in Fig.~\ref{F3}c the probability distribution
coincides exactly with $P_{abs}$, given by the line represented by circles.

The reason why this happens will be discussed in section~4.

\subsection{What is missing then? -- the $\delta$-function}

The difference between what we should get, see Eqs.(\ref{20},\ref{30}),
and what we have got, see Eqs.(\ref{60},\ref{70}), may be found in the
following simple formula

        \begin{equation}
S_{eff}=S_{abs}\pm i\pi\Theta(-\cos\theta).
        \label{80}
        \end{equation}
where $\Theta$ is the real step-function
        \begin{equation}
    \Theta(x)=\left\{
            \begin{array}{cl}
            1 &  x>0,\\
            0 &  x<0.
            \end{array}  \right.
        \label{80a}
        \end{equation}

{}From this action one gets the drift force
        \begin{equation}
-\frac{\delta S_{eff}}{\delta\theta}=-\frac{\delta S_{abs}}{\delta\theta}
\mp i\pi\sin\theta\delta(\cos\theta),
        \label{90}
        \end{equation}
and the corresponding Langevin equation
        \begin{equation}
\dot \theta=-\beta\sin\theta-P(\tan\theta)
\mp i\pi\sin\theta\delta(\cos\theta) +\eta,
        \label{91}
        \end{equation}
which is different from the normal one in Eq.(\ref{40}).

As was stated in the first paragraph of section 2.2, one will never get the
$1/\beta$-behaviour in case the complex distribution collapses to the real
one, independently of whether the segregation phenomenon is observed or
not. An important term missing in (\ref{40})
is an imaginary part in the drift to allow
for a complex distribution.

In this context, the additional drift force $i\pi\delta(-\cos\theta)$
in (\ref{90})
is expected to work well because it kicks the
configuration into the deep imaginary region.
This pushes the distribution collapsed on the real axis
to that in the whole complex plane and
may help to give the $1/\beta$ singularity.

\section{Theoretical support of the idea}

\setcounter{equation}{0}

In the previous section we have discussed how one may simulate correctly
a system whose drift force is singular.
Before starting to solve the new Langevin
equation obtained there,
we give some theoretical support for the idea of a $\delta$-function-like
drift.
Furthermore, an important question to be answered
is {\it ``whether our observation contradicts to the segregation
theorem or not?'' }.
To make the discussion clearer we consider an alternative model which is
simpler but has a similar structure as the
$\cos\theta$-model in (\ref{20}) and (\ref{30}).

\subsection {A simpler model (modified Rayleigh-model)}
Let us consider the following integral,
     \begin{equation}
<x>\equiv\frac{\int_{-\infty}^{+\infty} dx x^2
                        \exp(-\frac{\alpha}{2}(x-\beta)^2)}
                {\int_{-\infty}^{+\infty} dx x
           \exp(-\frac{\alpha}{2}(x-\beta)^2)}=\beta+\frac{1}{\alpha\beta}.
     \label{100}
     \end{equation}
One can rewrite it as
     \begin{equation}
<x>=\int_{-\infty}^{+\infty} dx x P_{eff}(x)
     =\frac{\int_{-\infty}^{+\infty} dx x \exp\{-S_{eff}\}}
           {\int_{-\infty}^{+\infty} dx \exp\{-S_{eff}\}},
     \label{110}
     \end{equation}
where
     \begin{equation}
S_{eff}=\frac{\alpha}{2}(x-\beta)^2-\ln x.
     \label{120}
     \end{equation}
As is seen in (\ref{100}), the quantity $<x>$ is also singular in the limit
of $\beta\to 0$. This singularity arises from the fact that the denominator
in (\ref{100}) vanishes in this limit although the numerator remains
finite. The situation is very similar to that of the $\cos\theta$-model.
Therefore, this model will be a good candidate to check all of the points
discussed in the previous sections.  Exactly in the same way as
in case of the $\cos\theta$-model, the naive Langevin equation\footnote
{
Note that this Langevin equation is closely related to the well known
Rayleigh process, $\dot x =-\alpha x +{\mu\over x}+\eta(t)$.
}
     \begin{equation}
\dot x =-\alpha (x-\beta)+{1\over x}+\eta(t).\
     \label{120a}
     \end{equation}
fails to reproduce the $1/\beta$ singularity given in (\ref{100}).
Based on this simple model, therefore, we will first study the
segregation theorem in more detail in the next subsection.

\subsection {The segregation phenomenon}

Let us start from the Fokker-Plank equation equivalent to
the above Langevin equation,
     \begin{eqnarray}
\dot P(x;t)&=&{1\over 2}{\partial^2\over\partial x^2}[B(x,t)P(x,t)]
     -{\partial\over\partial x} [A(x,t)P(x,t)],\nonumber\\
B(x,t)&=&2,\qquad A(x,t)=A(x)\equiv -\alpha (x-\beta)+{1\over x}.
     \label{120b}
     \end{eqnarray}
This is a second-order parabolic
differential equation. In order to specify its solution one needs an
initial condition, e.g., $P(x;t_0)=\delta(x-x_0)$, and boundary
conditions.
In case the drift force $A(x,t)$ is singular
(or the diffusion coefficient $B(x,t)$ vanishes) at some point
$x=x_b$, a certain boundary condition should be imposed at this point
which is called a {\it prescribed boundary condition}.
What kind of boundary condition one should take at $x=x_b$ is determined
by the form of the drift and the diffusion coefficient.
An extensive analysis of this problem has been given by
Feller\cite{FEL52}.
See also \cite{GAR83} for a review.

For our Langevin equation (\ref{120a}) the prescribed boundary condition
will be given at the singular point of the drift,
$x=x_b=0$.
$[x_b,x_0]$.
Following the classification of Feller \cite{FEL52}, the singular
point $x=x_b=0$ should be treated as an {\it entrance boundary}.
In this case the solution of the Langevin equation
(\ref{120a}) should be such that
if the configuration starting from, e.g., $x=x_0>x_b$ comes near to
that boundary it should be pushed back to the original direction $x>x_b$.
Therefore, the configuration starting from $x>x_b$
can never leave that region\footnote
{
When the configuration starts exactly
from that point $x=x_b$ it should enter either the region $x>x_b$
or $x>x_b$, which is the
origin of the name {\it entrance boundary}.
}.
This is the so called segregation phenomenon.

\subsection {The reason for the failure of the naive Langevin \newline
equationand the role of the $\delta$-function-like drift}

Similar to what we saw in Fig.~2 and Fig.~3, however, this is not the case
when one naively does {\it a numerical simulation} for solving the
Langevin equation
(\ref{120a}) {\it by discretizing the time $t$}. The configuration
turned out to pass through the barrier at $x=x_b$ easily and the segregation
theorem
looks to be violated.
The main reason why this happens may come from the fact that by
discretizing the time
we can not treat the singularity in the drift rigorously.
The adjusted time step has been introduced in order to
take
into account the effect of the singular region as rigorous as possible.
But this seems still not enough.
The system which are simulated by the discretized version
of the Langevin equation (\ref{120a}) can not exactly
be the process given by (\ref{120a}) but that with some regularized
drift $A(x)_R$.
In Figure 4, $A(x)_R$ is taken to be
constant $\pm 1/r_0$ for $|x|<r_0$.
The form of $A(x)_R$ in $|x|<r_0$ is, by no means, unique.
However, one can easily check that
the following discussion does not depend on the way how one
regularizes $A(x)$ in this region.
        \begin{figure}[p]
        \vspace*{5cm}
        \caption{}
         \label{F3.5}
         \begin{quotation}
(a) Regularized drift force $A(x)_R$ at $x\sim 0$.
(b) Solution of (\ref{120d}) under an assumption of continuous $P(x)$.
(c) Solution of (\ref{120d}) under an assumption of continuous
$\partial_x P(x)$.
         \end{quotation}
         \end{figure}

For the stochastic process with the above regularized drift $A(x)_R$,
no prescribed boundary condition exists by definition.
The solution, therefore, can pass through the point $x=0$ without
any problem. In addition to this,
an interesting point is that for this regularized process
the stationary solution for $P(x)$ (at $|x|>r_0$) is given
by $P_{abs}\propto|x|\exp(-{\alpha\over 2}(x-\beta)^2)$ and not by
$x\exp(-{\alpha\over 2}(x-\beta)^2)$.
To show this, let us try to solve the stationary solution $P(x)$ in the
case of the regularized drift,
     \begin{equation}
     \left ( {\partial\over\partial x} +A(x)_R \right )P(x)=0
     \label{120d}
     \end{equation}
The solution can be represented as (for simplicity we take $\beta=0$
in the following)
\begin{eqnarray}
  r_0<x   &:& P(x)=c_1 e^{-{\alpha\over 2}x^2+\ln x}, \nonumber\\
  0<x<r_0 &:& P(x)=c_2 e^{-{\alpha\over 2}x^2+{1\over r_0} x},\nonumber\\
  -r_0<x<0&:& P(x)=c_3 e^{-{\alpha\over 2}x^2-{1\over r_0} x},\\
  x<-r_0  &:& P(x)=c_4 e^{-{\alpha\over 2}x^2+ \ln x},\nonumber
   \label{sol}
\end{eqnarray}
with $c_1$,$c_2$,$c_3$ and $c_4$ being constants.
The continuity conditions of $P(x)$ at $x=\pm r_0$ and $x=0$ gives
an interesting relation between $c_1$ and $c_4$, namely
     \begin{equation}
c_4=c_1e^{\ln(-1)} = -c_1.
     \label{120e}
     \end{equation}
This proves that, outside of the singular region, the Fokker-Planck
distribution has the form of $P(x)\propto |x| \exp\{ (-\alpha/2)x^2\}$.
{}From those discussions, it may be clear why our numerical simulation
gave the results corresponding to $P_{abs}(\theta)$ given in
(\ref{60}) and in (\ref{70}) in the previous section.

Our proposal for the way out of this problem mentioned in the last
subsection is to include
an extra $\delta$-function type drift in the
Langevin equation
due to the relation $\ln x=\ln |x| \pm i\pi\Theta(-x)$.
The drift coming from $\ln |x|$, i.e.,
$-\delta (\ln |x|)/\delta x=P(1/x)$, is corresponding
in some sense to the regularized
drift $A(x)_R$ discussed above. Then the total drift should be\footnote
{
Note that an addition of this $\delta$-function type drift will never
spoil the regularity of the stochastic process at $x=0$. This can easily
be seen by calculating the functions
$f(x)$, $g(x)$, $h_1(x)$ and $h_2(x)$ defined
in \cite{GAR83} by using this drift. All of them turn out to be integrable
in the region of $[0,x_0]$ for any finite $x_0$.
}.
$\bar A(x)=A(x)_R+i\pi\delta(x)$
It is interesting to repeat the same discussion as above
using the new drift $\bar A(x)$. For this drift $\bar A(x)$, the
solution of $\bar J\equiv(\partial_ x +\bar A(x))P(x)=0$
can not, in general, be continuous at $x=0$. This is seen by
integrating both sides of this equation
in a small region $[-\epsilon,+\epsilon]$, and taking the
limit $\epsilon\to 0$. On the other hand
applying the same discussion to the stationary F-P equation,
$\dot P=\partial_x \bar J=\partial_x(\partial_x+\bar A(x))P=0$, one finds
that $\partial_x P(x)$ can still be continuous at $x=0$.
{}From this continuity condition
of $\partial_x P(x)$ one gets $c_4=c_1$ rather than that in (\ref{120e}).
This means that $P(x)\propto x\exp({-\alpha}x^2/2)$ rather than
$P(x)=P_{abs}(x)\propto|x|\exp{-\alpha x^2/2}$.

This shows that if we succeed to estimate the effect
of the $\delta$-function type drift in the Langevin equation, we will get
the correct result.

\section {How can one do the simulation for the $\delta$-function
 type drift?}

\setcounter{equation}{0}

\subsection{A model with a real $\delta$ function}

In order to study how to solve the Langevin equations with the
$\delta$-function type drift numerically, we first consider an even simpler
model given by the real action
     \begin{equation}
S_2=\frac{\alpha}{2}(x-\beta)^2-\frac{c}{2}\epsilon(x),
     \label{130}
     \end{equation}
where $c>0$ is a real constant. ue to the existence of the
$-c\epsilon(x)/2$ in the action, the distribution $P_2\propto
e^{-S_2}$ has a discontinuity at $x=0$, see Fig.~5.
The probability distribution is
shifted to the direction of positive $x$, which increases the value
of, e.g., $<x>$.

        \begin{figure}[p]
        \vspace*{5cm}
        \caption{}
         \label{F4}
         \begin{quotation}
Schematic figure of the probability distribution $P_2\propto e^{-S_2}$
with $S_2$ given in (\ref{130}).
         \end{quotation}
         \end{figure}

In the Langevin simulation, this effect will be given by the
drift $-\delta/\delta x (-c/2\epsilon(x))=c\delta(x)$ in the
corresponding Langevin equation
     \begin{equation}
\dot x=-\alpha (x-\beta)+c\delta(x)+\eta
     \label{140}
     \end{equation}
When the configuration crosses $x=0$ while updating the Langevin equation, it
gets a kick towards positive $x$, which works to increase the
expectation value $<x>$.

\subsection{A success of the numerical simulation}

For the numerical solution of the Langevin equation we first have to fix
how to  treat the drift $c\delta(x)$. In this subsection we like to discuss
two ways,
\begin{description}
\item[--]  the use of a smeared $\delta$-function, and
\item[--]  an integration of the $\delta$-function around $x=0$.
\end{description}

\noindent{\bf The use of the smeared $\delta$-function:}\qquad
One can replace the $\delta$-function by the smeared one,
     \begin{equation}
\delta(x)\longrightarrow \delta_{\epsilon}\;, \qquad
     \delta_\epsilon\equiv\frac{1}{\pi}\frac{\epsilon}{x^2+\epsilon^2},
     \label{150}
     \end{equation}
where $\epsilon$ is a small number.
In Fig.6, the  results for $<x>$ and $<x^2>$ simulated by the
Langevin equation (\ref{140}) with $\delta(x)$ being replaced
by the smeared one $\delta_\epsilon(x)$ are given. One can see
that, within this real model, the effect of the discontinuity in
the probability distribution is nicely simulated by the effect
of the $\delta$-function type drift in the Langevin equation.

        \begin{figure}[p]
        \vspace*{5cm}
        \caption{}
         \label{F6}
         \begin{quotation}
The result of $<x>$ and $<x^2>$ (for $\alpha =0.2, \beta = 0$) simulated
by the above Langevin algorithm versus $c$. For each run $50000$ iterations
have been done with $\Delta t=0.04$.  Errors are estimated from $32$ runs
of this kind. The solid line represents the exact result.
         \end{quotation}
         \end{figure}

\noindent{\bf Integrating out the $\delta$-function}:\qquad
It is also possible to get an effective algorithm by analytically
integrating out the $\delta$-function.
Let us rewrite the Langevin equation (\ref{140}) in a discretized form,
and consider the step from $x_i$ to $x_f$ ($\Delta x\equiv x_f-x_i$),
     \begin{eqnarray}
\Delta x&=&a+c\frac{\Delta\Theta(x)}{\Delta x}\Delta t, \nonumber \\
   a  &\equiv& \left( -\alpha (x-\beta)+\sqrt{\frac{2}{\Delta t}}\xi
                                       \right)\Delta t,
     \label{160}
     \end{eqnarray}
where $\xi$ is a Gaussian noise with $<\xi^2>=1$.
When $x$ does not cross $x=0$ within this step, $\Delta \Theta(x)=0$
and therefore from (\ref{160}) we simply get
     \begin{equation}
   \Delta x=a
\label{161}
     \end{equation}
When $x$ crosses $x=0$, $\Delta\Theta(x)=\pm 1$ ( depending on whether
$x_i>0$ or $x_i<0$ ) and the Langevin equation
(\ref{160}) becomes a second order algebraic equation for $\Delta x$.
Solving this equation we get
     \begin{eqnarray}
   \Delta x&=&\frac{a+\sqrt{a^2+4c\Delta t}}{2} \qquad(x_i<0,a>0),
                                                        \nonumber \\
   \Delta x&=&\frac{a-\sqrt{a^2-4c\Delta t}}{2} \qquad(x_i>0,a<0).
\label{162}
\end{eqnarray}
The simulated results for $<x>$ by the use of this algorithm is shown
in Fig.~\ref{F7}. One can see that the Langevin simulation with the
$\delta$-function type drift is being able to recover the exact one
very nicely.

        \begin{figure}[p]
        \vspace*{5cm}
        \caption{}
         \label{F7}
         \begin{quotation}
The result of $<x>$ and $<x^2>$ (for $\alpha =0.2, \beta = 0$) simulated
by the above Langevin algorithm versus $c$. For each run $50000$ iterations
have been done with $\Delta t=0.01$.  Errors are estimated from $32$ runs
of this kind. The solid line represents the exact result.
         \end{quotation}
         \end{figure}

\section {Complex $\delta$-function}

\setcounter{equation}{0}

In this section we come back to our original problem, the
$\cos\theta$-model.
As discussed in section 3.2 the Langevin equation for
this system is given by (\ref{91}). This Langevin equation has an
additional
drift $i\pi\delta(\cos\theta)$ that kicks the configuration $\theta$ into the
deep imaginary region. This will surely help to give a
big $<\cos\theta>$ because of the reasons explained in section 2.2.
The problem, however, is how to continue this $\delta$-function into the
whole complex $\theta$-plane.

In order to explain the situation more clearly, let us once more use the
modified Rayleigh model given in (\ref{100}). The Langevin equation for
this system is given by (\ref{120a}) with the
additional drift $i\pi\delta(x)$. In this case, even when one starts
the iteration from some point on the real axis, it
becomes complex after getting a kick by the $\delta$-function.
Then the question arises how to treat the $i\pi\delta(x)$ in the
complex plane $z=x+iy$.

\noindent
{\bf Smeared $\delta$-function: \qquad} One possibility may be to use the
smeared $\delta$-function continued into the complex plane,
i.e., $\delta_\epsilon(z)=\epsilon/\pi(z^2+\epsilon^2)$.
But this does not work. The reason may be that
the above smeared $\delta$ function
which has been introduced in order to
take into account the effect of the singularity in the drift $1/x$
has again a new
singularity at $z=\pm i\epsilon$ in the complex plane.
Actually we have tried to use it for simulation, it could not
reproduce a correct $1/\beta$ behaviour for $<\cos\theta>$.

\noindent
{\bf Integrating out the $\delta$-function: \qquad} We also have a
problem in applying this idea to the complex case, because we do not have
the formula  corresponding to (\ref{160}) in the complex case. But neglecting
this fundamental question, we may use the solution of this equation,
i.e., (\ref{161}) and (\ref{162}), with $c=i\pi$. The result
is not good enough to produce the $1/\beta$ singularity although
it refines the results obtained without the
$\delta$-function-type drift.

In spite of these failure, it is still very encouraging
that with the idea of the $\delta$-function type kick
we can find an effective algorithm by which the above $1/\beta$
behaviour in  the $\cos\theta$ model is successfully simulated.
Let us approximate the drift $i\pi\sin\theta\delta(\cos\theta)$
in the Langevin equation (\ref{91}) in the following way \cite{SCH91}: Consider
the
region ${\cal D}$ in the complex plane of $\theta=\theta_r+i\theta_i$
around  the singular points $\theta=\pi/2$ or
$3\pi/2$,
     \begin{equation}
\theta\in {\cal D} \Longleftrightarrow |\Re\cos\theta|<\hat\delta,
\label{singreg}
     \end{equation}
$\hat\delta$ being a small constant, and take
     \begin{equation}
i\pi\sin\theta\delta(\cos\theta) \longrightarrow
      \left\{
            \begin{array}{rl}
               0,    &\quad\mbox{for $\theta\notin {\cal D}$}\\
               i\pi, &\quad\mbox{for $\theta\in {\cal D}$}
            \end{array}
      \right.
\label{ipikick}
     \end{equation}
In Fig.~\ref{F9} we show
the result of $<\cos\theta>$ for the $\cos\theta$ model.
The data drawn by black squares are those
simulated by the use of the Langevin equation (\ref{91}), where the
drift proportional to $\delta(\cos\theta)$ has been replaced by
the kick given by equation (\ref{ipikick}) above. One can see
that the simulated result coincides with the theoretical
prediction quite nicely in the whole region of $\beta$.

        \begin{figure}[p]
        \vspace*{5cm}
        \caption{}
         \label{F9}
         \begin{quotation}
The black square and the circle represents, respectively, the result
of $<\cos\theta>$ with and without the kick (\ref{ipikick}). For the parameters
$\delta=0.07$,$\Delta t=0.002$ have been taken. In order to calculate the
average $100000$ iterations has been done for each run and the error has
been estimated from $8$ different kind of runs.
The solid line represents the theoretical prediction.
Dashed line and dotted line corresponds, respectively, to the theoretical
prediction $<\cos\theta>_1$ and $<\cos\theta>_{abs}$ explained in
subsection 2.2 and section 3.

         \end{quotation}
         \end{figure}

In spite of this wonderful success of the numerical simulation, we
have not yet found any smart theoretical justification of the above
algorithm. Moreover the effectiveness of this algorithm can not
be universal for other models.  The modified Rayleigh model discussed in
section 4 is a good example. If the $\delta$-function type drift
$i\pi\delta(x)$ for this model is treated exactly
in the same way as above, it will just give the imaginary kick $i\pi$
to the configurations near to the singular point $z=0$. The big imaginary
part in $z$, however, does not directly give a big $\Re<z>=<x>$ contrary
to the case of the $\cos\theta$ model.
We need therefore deeper understanding of the algorithm
found above from more theoretical view point like that in the case of
those models with real $\delta$-function in the previous section.

\section{Conclusion}

\setcounter{equation}{0}

In the numerical simulation of certain field theoretical models,
the complex Langevin simulation
has been believed to fail due to the violation of the ergodicity.
In this paper we have given a detailed analysis of this problem
based on a toy model in one degree of freedom
($\cos\theta$ model).
The corresponding Langevin equations
involved in the above problem have a singular drifts, e.g.,
$\;\propto \tan\theta$.
Our observation is that the failure is not due to the
defect of the complex Langevin simulation itself, but rather to he way how one
treats
the above singular drift force.
We  have tried to justify this statement using also some alternative models.
Under the above observation we could also give an effective algorithm by
which we can simulate
wonderfully the ${1/\beta}$ behaviour of the expectation value
$<\cos\theta>$ in the limit of $\beta\to 0$.

Unfortunately, however, the final theoretical justification of the
above effective algorithm is still missing. When one succeeds in
getting the rigorous theoretical background about how to treat
the singular drift in the complex Langevin simulation, practical
gain in the numerical simulation of lattice field theory
will be extremely large.

\vskip 1truecm

\noindent{\bf Acknowledgment}

One of the authors (K.O.) acknowledges financial support
from the Deutscher Akademischer Austauschdienst (DAAD)
for his stay in Germany, during when a part of this paper has
been completed. The authors also like to acknowledge financial support from the
Research Institute for General Economy, Tokuyama University.


\end{document}